\documentclass[12pt]{article}%
\usepackage{amsmath}%
\usepackage{txfonts}%
\usepackage{amsfonts}%
\usepackage{amssymb}%
\usepackage{mathrsfs}
\usepackage{graphicx}
\usepackage{color}
\usepackage{setspace}
\usepackage{esint}
\usepackage[rightcaption]{sidecap}
\usepackage{wrapfig}
\usepackage{lipsum}
\usepackage{caption}
\usepackage{longtable}
\usepackage[T2A,T1]{fontenc}
\usepackage[utf8]{inputenc}
\usepackage[russian,english]{babel}
\usepackage[toc,page]{appendix}

\numberwithin{equation}{section}

\newcommand{\rr}{\color{red}}

\newcommand{\be}{\begin{equation}}
\newcommand{\ee}{\end{equation}}

\usepackage{anysize}
\marginsize{2.5cm}{2.5cm}{1cm}{2cm}
\usepackage{mathptmx}
-1in\relax 
\begin{document}
\title{Ramjet Acceleration of Microscopic Black Holes Within Stellar Material}
\maketitle
\begin{center}
\begin{tabular}{l}
%Marianna A. Shubov$^*$            &
\hskip1cm Mikhail V. Shubov         \\
%University of New Hampshire       &
\hskip1cm University of MA Lowell    \\
%33 Academic Way                   &
\hskip1cm One University Ave,        \\
%Durham, NH 03824                  &
\hskip1cm Lowell, MA 01854           \\
%E-mail: marianna.shubov@gmail.com &
\hskip1cm E-mail: mvs5763@yahoo.com  \\
\end{tabular}
\end{center}
%\hskip1.7cm*\emph{Corresponding Author.}  E-mail: marianna.shubov@gmail.com

\normalsize
\begin{center}
  \textbf{Abstract.}
\end{center}
In this work we present a case that Microscopic Black Holes (MBH) of mass $10^{16}\ kg$ to $3 \cdot 10^{19}\ kg$ experience acceleration as they move within stellar material at low velocities.  The accelerating forces are caused by the fact that a MBH moving through stellar material leaves a trail of hot rarified gas.  The rarified gas behind a MBH exerts lower gravitational force on the MBH than the dense gas in front of it.  The accelerating forces exceed the gravitational drag forces when MBH moves at Mach number $\mathcal{M}<\mathcal{M}_0$.  The equilibrium Mach number
$\mathcal{M}_0$ depends on MBH mass and stellar material characteristics.  Our calculations open the possibility of MBH orbiting within stars including Sun at Mach number $\mathcal{M}_0$.  At the end of this work we list some unresolved problems which result from our calculations.

\section{Introduction.}

In the research presented in the works \cite{Earth,NStar,Carr}, it has been suggested that Primordial Black Holes make up a significant fraction of dark matter.  Microscopic Black Holes [MBH] can also be formed within stars by coalescence of dark matter composed of weakly interacting massive particles \cite{Type1a,wimp1}.

Up to now, researchers believed that all MBH captured by a star would be slowed down within stellar material until they settle in the stellar center \cite{Earth,NStar}.
In the present work, we explore the possibility of MBH accelerating during their passage through matter.  As MBH passes through matter, it accretes material at a rate we denote $\dot{M}$ and generates energy by accretion.  Most of the accreted mass is absorbed by the MBH, while about 0.5\% of the mass is turned into the energy of gamma and proton radiation.  This radiation heats and rarefies the surrounding material.  Dense material ahead of moving MBH exerts greater gravitational pull on the moving MBH than the rarefied material behind it.  \emph{As a result, the moving MBH experiences a net forward force.}  We call this force the \textbf{MBH ramjet} force.  The effect is illustrated below on Fig. 1.
\begin{center}
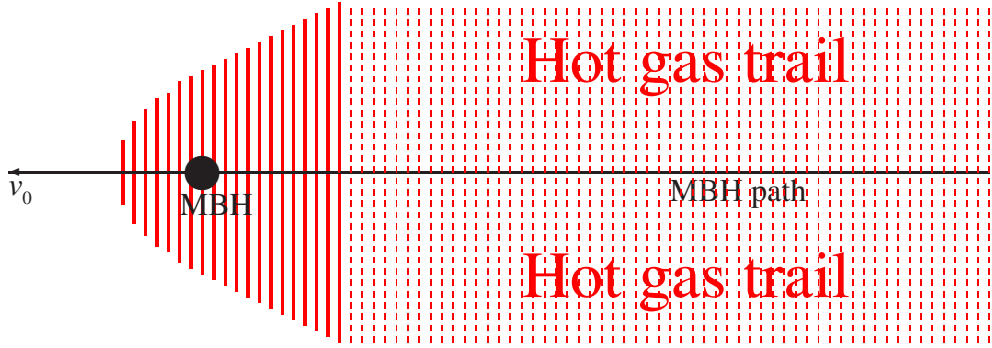

\setlength{\unitlength}{1.5mm}
\begin{picture}(90, 30)
\linethickness{0.3mm}
%\qbezier(29,30)(-10,15)(29,0)
\multiput(29,0)(-1,.5){15}{\rr{\line(0,1){15}}}
\multiput(29,30)(-1,-.5){15}{\rr{\line(0,-1){15}}}
\put(14,8){\rr{\line(0,1){14}}}
\put(13,8.5){\rr{\line(0,1){13}}}
\put(12,9.4){\rr{\line(0,1){11.2}}}
\put(11,10.5){\rr{\line(0,1){9}}}
\put(10,12.2){\rr{\line(0,1){5.6}}}
\put(17,15){\circle*{3}}
\put(10,15){\vector(-1,0){10}}
\put(15,11.2){MBH}
\put(0,13){$v_{_0}$}
\put(3,15){\line(1,0){83}}
\put(58,12.5){MBH path}
\linethickness{0.1mm}
\put(45,23){\Huge \rr{Hot gas trail}}
\put(45,4){\Huge \rr{Hot gas trail}}
\multiput(29,0)(1,0){58}{\rr{\multiput(0,0)(0,1){30}{\line(0,1){.5}}}}
\linethickness{0.2mm}
\end{picture}
\captionof{figure}{MBH passage through matter. \label{F01}}
\end{center}

In this work we derive the conditions under which MBH accelerates within the stellar material.  In order to express these conditions mathematically, we define three efficiencies (gas redistribution efficiency, radiative efficiency, accretion efficiency) involved in the MBH passage through stellar material.  \textbf{Gas redistribution efficiency}, $\eta_{_G}$, is the ratio of the accelerating force caused by gas rarefication behind the MBH to theoretical maximum of such force.  The exact definition starts at paragraph containing (\ref{3.07}) and ends with paragraph containing (\ref{3.09}).  \textbf{Radiative efficiency}, $\eta_{_{\Gamma}}$, is the ratio of the total power radiated by MBH to the energy $\big( M c^2 \big)$ of the mass falling into MBH.  It is expressed in (\ref{3.10}).  \textbf{Accretion efficiency}, $\eta_{_A}$, is the ratio of the actual and the zero-radiation mass capture rates.  It is defined in (\ref{3.14}).

We show that in case of MBH moving through stellar material at supersonic speed, the condition for MBH acceleration is given in (\ref{3.28}):
\be
\label{1.01}
\mathcal{N}=\frac{\eta_{_A} \eta_{_{\Gamma}} \eta_{_G}^2}{T_6}
\gtrsim  4 \cdot 10^{-4},
\ee
where $T_6$ is the temperature of the stellar material in millions Kelvin.  Even though we do not have precise values for efficiencies $\eta_{_A}$, $\eta_{_{\Gamma}}$, and $\eta_{_G}$, we are almost certain that for supersonic MBH, condition (\ref{1.01}) is never met.  In case of MBH moving through stellar material at a subsonic speed, the condition for MBH acceleration is given in (\ref{3.32}):
\be
\label{1.02}
\mathcal{N}=\frac{\eta_{_A} \eta_{_{\Gamma}} \eta_{_G}^2}{T_6}
\gtrsim  9 \cdot 10^{-7} \mathcal{M}^3 \mathfrak{F}\big(\mathcal{M},\eta_{_A}\big),
\ee
where $\mathcal{M}$ is the Mach number and
$\mathfrak{F}\big(\mathcal{M},\eta_{_A}\big) \in (0.11,1)$ is given in (\ref{3.33}).  MBH moving with supersonic speed always experience deceleration within stellar material.
MBH moving at subsonic speed experiences acceleration when the Mach number exceeds $\mathcal{M}_0$ (the equilibrium Mach number) and deceleration when the Mach number is below $\mathcal{M}_0$.  Eventually the MBH settles into an intrastellar orbit with Mach number $\mathcal{M}_0$.  The value of $\mathcal{M}_0$ can be obtained by solving (\ref{1.02}) as an equality.

In Appendix A, estimates from below and from above for $\eta_{_G}$ are calculated.  Obtaining more precise results remains an open problem.  Calculating the values of $\eta_{_A}$ and $\eta_{_{\Gamma}}$ also remain open problems.  As we discuss later in this work, different theorists obtained different results for $\eta_{_{\Gamma}}$.

We briefly outline the content of the present paper.
In Section 2, we present dimensionless representation as well as common ranges of mass, density, temperature, and speed.
In Section 3, we calculate forces acting on MBH.  We also derive conditions for MBH acceleration at subsonic and supersonic speed.
In Section 4, we present estimates for $\eta_{_A}$ and $\eta_{_{\Gamma}}$.
In Section 5, we present an empirical discussion of possible behaviors of MBH within stellar material.  We also discuss the possibility of MBH triggering Type 1a Supernovas.  In Section 6, the problems remaining after this work are briefly described.

\section{Dimensionless representation of mass, density, temperature, and speed}
In this section, we present the MBH mass and speed, as well as stellar material density and temperature ranges which may appear in the Universe.  We also present dimensionless representations of these quantities for simplicity.

According to the plot in \cite[p.14]{Carr}, considerations other than stellar capture constrain the masses of MBH as dark matter to the range of $10^{16}\ kg$ to
$5 \cdot 10^{21}\ kg$.
If $M_{_{\text{MBH}}}$ is the notation for the mass of MBH, then by $M_{18}$ we denote the ``normalized" mass defined by
\be
\label{2.01}
M_{18}=\frac{M_{_{\text{MBH}}}}{10^{18}\ kg} \in \big( 10^{-2}, 5 \cdot 10^3\big).
\ee

The characteristics of matter through which MBH passes are also of great diversity.  In this work, we consider a passage of MBH through stellar material composed primarily of ionized hydrogen and helium.  The MBH passes through gases of density $10^{-7}\ kg/m^3$ to $2 \cdot 10^5\ kg/m^3$.   The portions of MBH path where the gas density is below $10\ kg/m^3$ will not have significant effect on the MBH trajectory.
If $\rho$ is the density of material through which the MBH passes, then by $\rho_3$ we denote the ``normalized" (or medium) density defined by
\be
\label{2.02}
\rho_3=\frac{\rho}{10^3\ kg/m^3} \in \big( 10^{-2}, 2 \cdot 10^2 \big).
\ee
Notice that $\rho_3=1$ is water density.

Gas medium temperature is between $2 \cdot 10^6\ ^oK$ and $3 \cdot 10^7\ ^oK$.
If $T$ is the temperature of the surrounding material, then by $T_6$ we denote the ``normalized" temperature defined by
\be
\label{2.03}
T_6=\frac{T}{10^6\ ^oK} \in \big( 2, 30 \big).
\ee

The MBH speed within a star varies between $5 \cdot 10^5\ m/s$ and $2 \cdot 10^6\ m/s$.
If $v_0$ is the speed of MBH, then by $v_6$ we denote the ``normalized" speed
\be
\label{2.04}
v_6=\frac{v_0}{10^6\ m/s} \in \big( 0.5, 2 \big).
\ee

\section{Forces acting on a MBH passing through matter}
\subsection{Total force acting on a MBH}
There are three forces acting on a black hole passing through matter.
First, there is a decelerating force, denoted by $F_t$, due to the tidal drag or the gravitational drag.  For a MBH moving at a supersonic speed, $F_t$ is given by  \cite[p.8]{tidal}
      \be
      \label{3.01}
      F_t=-\frac{1}{v_0} P_t=-\frac{4 \pi (MG)^2 \rho}{v_0^2}
      \ln \left(\frac{r_{\max}}{r_{\min}}\right),
      \ee
where $P_t$ is the decelerating power produced by the drag force, $\rho$ is the density of the surrounding medium, $r_{\max}$ is the approximate distance from the MBH to the location where the stellar material is consistent, and $r_{\min}$ is the radius at which matter is initially unperturbed by the MBH radiation.  In formula (\ref{3.01}), $M$ is the mass of the MBH and $G$ is the gravitational constant.

We take $r_{\max}$ to be about $5 \cdot 10^7\ m$ for a Sun-like star.  We take $r_{\min}$ to be about 0.1 $m$.
Hence,
      \be
      \label{3.02}
      \ln \left(\frac{r_{\max}}{r_{\min}}\right) \approx 20.
      \ee
For an MBH travelling at Mach number $\mathcal{M} \lesssim 0.8$, the gravitational drag can be given by the following formula (see \cite[p.5]{tidalow}, \cite[p.69]{tidalow1}, \cite[p.8]{tidalow2})
      \be
      \label{3.03}
      F_t=-\frac{4 \pi (MG)^2 \rho}{v_0^2}
      \left[
      \frac{1}{2} \ln \left(\frac{1+\mathcal{M}}{1-\mathcal{M}} \right)
      -\mathcal{M}  \right].
      \ee
Since $\mathcal{M}<1$, equation (\ref{3.03}) can be rewritten in the form of a converging series
      \be
      \label{3.04}
      F_t=-\frac{4 \pi (MG)^2 \rho}{v_0^2}
      \sum_{n=1}^{\infty} \frac{\mathcal{M}^{2n+1}}{2n+1}.
      \ee

Second, there is a decelerating force due to mass acquisition.  As the MBH acquires mass, which is initially at rest, the MBH momentum does not change.  From conservation of momentum, we calculate the rate of change of MBH speed:
      \be
      \label{3.05}
      \frac{\partial p}{\partial t}
      =\frac{\partial}{\partial t} \big( M v_0 \big)
      =\dot{M} v_0+ M \dot{v}_0=0
      \qquad
      \Rightarrow
      \qquad
      \dot{v}_0= - v_0 \frac{\dot{M}}{M}.
      \ee
Using MBH speed change, we calculate the effective force as
      \be
      \label{3.06}
      F_m=M \frac{\partial v_0}{\partial t}=-\dot{M} v_0.
      \ee

It will be shown (see (\ref{3.22}) below) that for supersonic MBH, this force contributes less than 10\% to the total decelerating force.  For subsonic MBH, $F_m$ is the main decelerating force.

Third, there is an accelerating force due to matter rarefaction behind the moving MBH.  This force is denoted by $F_r$.  This force is estimated in Appendix A.

To define gas redistribution efficiency (see (\ref{3.09}) below), we need two radii, $r_1$ and $r_2$.
Define the radius $r_1$ in terms of the accelerating force $F_r$.
The radius $r_1$ is such that the gravitational effect of the rarefied gas is the same as that of a sphere of gas directly behind the MBH having radius $r_1$, and density $\rho/2$.  Then the relationship between $r_1$ and the  accelerating ``ramjet" force $F_r$ acting on the MBH is
      \be
      \label{3.07}
      F_r=\frac{M M_s G}{r_1^2}=M G \frac{\frac{1}{2} \frac{4}{3} \pi r_1^3 \rho }{r_1^2}=\frac{2}{3} \pi M G \rho r_1,
      \ee
where $M_s$ is the mass of the sphere of rarefied gas of density $\rho/2$.

The radius $r_2$ defined in terms of the power $P$ radiated by MBH passing through the stellar material.  Imagine an infinite cylinder filled with the stellar material and assume that a circular cross section of this cylinder is moving with speed $v_0$ so that its center slides along the cylinder axis.  Assume that all this power $P$ is spent to heat the gas on the trail of this moving cross section.  This would lead to an increase of the gas temperature.  We define $r_2$ as such radius of the cylinder that the temperature of gas in the above imaginary experiment would double.
\be
\label{3.08}
\begin{split}
P=\big(\text{Mass heated per unit of time} \big) \cdot T \cdot C_v
=v_0 \big(\pi r_2^2 \rho \big) T \cdot C_v= \pi v_0 \rho T C_v r_2^2,
\end{split}
\ee
where $C_v$ is the heat capacity of gas of stellar material at constant volume.

The \textbf{gas redistribution efficiency} is defined as
\be
\label{3.09}
\begin{split}
\eta_{_G}=\frac{r_1}{r_2}.
\end{split}
\ee
As we show later in this section, the ramjet force $F_r$ acting on MBH is proportional to $\eta_{_G}$. Calculation of $\eta_{_G}$ is complicated.  In Appendix A we perform some steps in calculating $\eta_{_G}$ for different MBH velocities and list the unsolved problems.
%it is shown that $\eta_{_G}>0.3$ for an MBH passing through stellar material at subsonic speed, and $\eta_{_G}<1.06$ for any MBH speed. Notice, that the sound velocity is determined with respect to the stellar material.

The radiative power of the MBH passing through stellar material is
\be
\label{3.10}
\begin{split}
P=\eta_{_{\Gamma}} c^2 \dot{M},
\end{split}
\ee
where $\eta_{_{\Gamma}}$ is the \textbf{radiative efficiency} of MBH and $\dot{M}$ is the mass accretion rate.  %The rate $\dot{M}$ is estimated below.
For an MBH moving through medium at supersonic speed, the Bondi-Hoyle-Lyttleton accretion rate is \cite[p.203]{Bondi}
\be
\label{3.11}
\begin{split}
\dot{M}_{BH} = 4 \pi r_b^2 \rho \sqrt{v_0^2+v_s^2}\ ,
\end{split}
\ee
where $r_b$ is the Bondi radius, $\rho$ is the stellar material density, $v_s$ is the sound speed in the stellar material, and $v_0$ is the MBH speed relative to the stellar material.
The Bondi radius is \cite[p.203]{Bondi}
\be
\label{3.12}
r_b=\frac{MG}{v_s^2+v_0^2}.
\ee
Substituting (\ref{3.12}) into (\ref{3.11}), we obtain
\be
\label{3.13}
\begin{split}
\dot{M}_{BH}
= 4 \pi r_b^2 \rho v_0
=\frac{4 \pi (MG)^2 \rho}{\left( v_0^2+v_s^2 \right)^{3/2}}.
\end{split}
\ee

Actual mass capture rate is considerably smaller.  The radiative heating of the gas surrounding MBH increases its' temperature.  This  increases the gas sound speed and decreases gas density.   Thus, the actual mass capture rate is
\be
\label{3.14}
\begin{split}
\dot{M}
\approx
\frac{4 \pi (MG)^2 \rho_r}{\left( v_0^2+v_{sr}^2 \right)^{3/2}},
\end{split}
\ee
where $v_{sr}$ is the sound speed at the accretion radius and $\rho_r$ is the density at the accretion radius.
 Recall the \textbf{accretion efficiency} $\eta_{_A}$ is the quotient of actual and zero-radiation mass capture rates:
\be
\label{3.15}
\eta_{_A}
=\frac{\dot{M}}{\dot{M}_{BH}}
\approx \left(\frac{v_0^2+v_s^2}{v_0^2+v_{sr}^2}\right)^{3/2} \frac{\rho_r}{\rho}.
\ee

Equating the power from (\ref{3.08}) and (\ref{3.10}), we obtain
\be
\label{3.16}
\begin{split}
\pi v_0 \rho T C_v r_2^2=\eta_{_{\Gamma}} c^2 \dot{M}.
\end{split}
\ee
Substituting (\ref{3.13}) and (\ref{3.14}) into (\ref{3.16}), we obtain
\be
\label{3.17}
\begin{split}
\pi v_0 \rho T C_v r_2^2=\eta_{_{\Gamma}} c^2 \eta_{_A}
\frac{4 \pi (MG)^2 \rho}{\left( v_0^2+v_s^2 \right)^{3/2}}.
\end{split}
\ee
Thus,
\be
\label{3.18}
\begin{split}
r_2=2 \sqrt{\eta_{_A}} \sqrt{\frac{\eta_{_{\Gamma}} c^2}{T C_v}}\
\frac{MG }{ \left( v_0^2+v_s^2 \right)^{3/4} \sqrt{v_0}}
=2 \sqrt{ \frac{\eta_{_A} \eta_{_{\Gamma}} c^2}{T C_v}}\
\frac{MG }{ v_0^2} \left(1+\frac{v_s^2}{v_0^2} \right)^{-3/4}.
\end{split}
\ee

Substituting (\ref{3.18}) into (\ref{3.09}), we obtain an expression for $r_1$:
\be
\label{3.19}
\begin{split}
r_1=2 \eta_{_G} \sqrt{\frac{\eta_{_A} \eta_{_{\Gamma}} c^2}{T C_v}}\
  \frac{MG}{v_0^2} \left(1+\frac{v_s^2}{v_0^2} \right)^{-3/4}.
\end{split}
\ee

At this point we calculate the second and the third forces acting on the MBH.
The first one is given in  (\ref{3.01}) for an MBH moving through stellar material at supersonic speed and in (\ref{3.03}) for an MBH moving through stellar material at subsonic speed.
Substituting (\ref{3.19}) into (\ref{3.07}), we obtain
\be
\label{3.20}
F_r
=\frac{2}{3} \pi M G \rho r_1
=\left[
\frac{4}{3} \eta_{_G} \sqrt{\frac{\eta_{_A} \eta_{_{\Gamma}} c^2}{T C_v}} \left(1+\frac{v_s^2}{v_0^2} \right)^{-3/4}
\right]
\frac{\pi (MG)^2 \rho}{v_0^2}.
\ee
Substituting (\ref{3.13}) and (\ref{3.14}) into (\ref{3.06}), we obtain
\be
\label{3.21}
\begin{split}
F_m&=-\dot{M} v_0
=-\eta_{_A} \dot{M}_{BH} v_0
=-\eta_{_A}\ \ \frac{4 \pi (MG)^2 \rho}{\left( v_0^2+v_s^2 \right)^{3/2}}
\ \ v_0=
-4 \eta_{_A} \left(1+\frac{v_s^2}{v_0^2} \right)^{-3/2} \frac{\pi (MG)^2 \rho}{v_0^2}.
%-\eta_{_A} \frac{4 \pi (MG)^2 \rho}{v_0}
\end{split}
\ee
\subsection{Conditions for supersonic MBH acceleration}
If the MBH travels through stellar material at supersonic speed with Mach number $\mathcal{M} \gtrsim 1.5$, then the tidal friction force acting on MBH is given by (\ref{3.01}).  The total force acting on MBH is obtained by summing (\ref{3.01}), (\ref{3.20}), and (\ref{3.21}) and it can be given by the following formula:
\be
\label{3.22}
\begin{split}
F&=F_t+F_m+F_r\\
&=\frac{\pi (MG)^2 \rho}{v_0^2}\left[
-4 \ln \left(\frac{r_{\max}}{r_{\min}}\right)
-4 \eta_{_A}\left(1+\frac{v_s^2}{v_0^2} \right)^{-3/2}
+2 \eta_{_G} \sqrt{\frac{\eta_{_A} \eta_{_{\Gamma}} c^2}{T C_v}}
\left(1+\frac{v_s^2}{v_0^2} \right)^{-3/4}
\right].
\end{split}
\ee

The above equation shows that MBH will accelerate if and only if $F>0$, i.e.
\be
\label{3.23}
\eta_{_G} \sqrt{\frac{\eta_{_A}\eta_{_{\Gamma}} c^2}{T C_v}}
\left(1+\frac{v_s^2}{v_0^2} \right)^{-3/4}>
2 \ln \left(\frac{r_{\max}}{r_{\min}}\right)
+2 \eta_{_A} \left(1+\frac{v_s^2}{v_0^2} \right)^{-3/2} .
\ee

In this subsection we estimate conditions under which the MBH passing through matter accelerates, i.e., (\ref{3.23}) holds.  This condition can be rewritten as
\be
\label{3.24}
\eta_{_A} \eta_{_{\Gamma}} \eta_{_G}^2 >
\frac{4 T C_v}{c^2}
\Bigg[\ln \left(\frac{r_{\max}}{r_{\min}}\right)
+ \eta_{_A} \left(1+\frac{v_s^2}{v_0^2} \right)^{-3/2} \Bigg]^2
\left(1+\frac{v_s^2}{v_0^2} \right)^{3/2}
.
\ee
Based on solar density \cite[pp.378-379]{solar}, the velocity of an MBH coming into the center of a sun-like star from a high apogee elliptical orbit is about 1,390 $km/s$.  The sound velocity in the center of a sun-like star is 506 $km/s$ \cite[pp.378]{solar}.  Hence, for an MBH coming into the center of a sun-like star from a high apogee orbit, the Mach number is $\mathcal{M} \approx 2.7$, which yields
\be
\label{3.25}
\left(1+\frac{v_s^2}{v_0^2} \right)^{3/2} \approx 1.2.
\ee
Recalling (\ref{3.02}), and the fact that $\eta_{_A} < 1$, we rewrite the estimate to (\ref{3.24}) as (see (\ref{1.01})
\be
\label{3.26}
\eta_{_A} \eta_{_{\Gamma}} \eta_{_G}^2
\gtrsim \frac{T C_v}{c^2} \cdot 1.7 \cdot 10^3
= T_6\ \frac{10^6 C_v}{c^2} \cdot 1.7 \cdot 10^3.
\ee
 The heat capacity at constant volume of monatomic gas is
\be
\label{3.27}
C_v=\frac{3R}{2 m_a},
\ee
where $m_a$ is the average molar mass of the gas, and $R$ is the gas constant
($R=8.314\ \frac{J}{mol\  ^oK }$).  Typical stellar material consists of monatomic gas with average particle mass of $0.62\ amu$ \cite[p.378]{solar}.  Hence, the heat capacity at constant volume for stellar material is  $C_v=2.01 \cdot 10^4\ \frac{J}{kg\ ^oK}$.  Thus, (\ref{3.26}) can be rewritten as (see (\ref{1.01}))
\be
\label{3.28}
\mathcal{N}=\frac{\eta_{_A} \eta_{_{\Gamma}} \eta_{_G}^2}{T_6}
\gtrsim  4 \cdot 10^{-4}.
\ee
As we discuss in Subsection 4.3, different calculations of $\eta_{_{\Gamma}}$ yield different results, yet all of them are below $0.1$.  As we show in Subsection 4.2, $\eta_{_{\Gamma}}$ is very small if the temperature of gas at Bondi radius is high.  The solar gas temperature exceeds $T_6=4$ for radius under 0.5 Solar radii \cite{solar}.

%In Appendix A we prove that $\eta_{_G}<1.06$.

In the conclusion of this subsection, we can be almost certain that relation (\ref{3.28}) does not hold for Mach numbers $\mathcal{M}>1$, thus MBH can not accelerate in a supersonic flight regime.  In order to prove this assertion rigorously, we need extensive analysis which is not only beyond the scope of this work but also beyond the scope of any previous work on black hole accretion.

\subsection{Conditions for subsonic MBH acceleration}
If the MBH travels through stellar material at subsonic speed with Mach number $\mathcal{M} \lesssim 0.8$, then the tidal friction force acting on MBH is given by (\ref{3.04}).  The total force acting on MBH is obtained by summing (\ref{3.04}), (\ref{3.20}), and (\ref{3.21}), leading to
\be
\label{3.29}
\begin{split}
F&=F_t+F_m+F_r\\
&=\frac{\pi (MG)^2 \rho}{v_0^2}\left[
-4 \sum_{n=1}^{\infty} \frac{\mathcal{M}^{2n+1}}{2n+1}
-4 \eta_{_A}\left(1+\frac{v_s^2}{v_0^2} \right)^{-3/2}
+2 \eta_{_G} \sqrt{\frac{\eta_{_A} \eta_{_{\Gamma}} c^2}{T C_v}}
\left(1+\frac{v_s^2}{v_0^2} \right)^{-3/4}
\right]\\
&=\frac{\pi (MG)^2 \rho}{v_0^2}\left[
-4 \sum_{n=1}^{\infty} \frac{\mathcal{M}^{2n+1}}{2n+1}
-4 \eta_{_A}\Big(1+\mathcal{M}^{-2} \Big)^{-3/2}
+2 \eta_{_G} \sqrt{\frac{\eta_{_A} \eta_{_{\Gamma}} c^2}{T C_v}}
\Big(1+\mathcal{M}^{-2} \Big)^{-3/4}
\right].
\end{split}
\ee
The above equation shows that MBH will accelerate if and only if $F>0$ or
\be
\label{3.30}
\eta_{_G} \sqrt{\frac{\eta_{_A} \eta_{_{\Gamma}} c^2}{T C_v}}
\Big(1+\mathcal{M}^{-2} \Big)^{-3/4}
>
2 \sum_{n=1}^{\infty} \frac{\mathcal{M}^{2n+1}}{2n+1}
+2 \eta_{_A}\Big(1+\mathcal{M}^{-2} \Big)^{-3/2}.
\ee
Inequality (\ref{3.30}) can be rewritten as the following estimate:
\be
\label{3.31}
\begin{split}
\eta_{_A} \eta_{_{\Gamma}} \eta_{_G}^2 &>
\frac{4 T C_v}{c^2} \Big(1+\mathcal{M}^{-2} \Big)^{3/2}
\left[
\sum_{n=1}^{\infty} \frac{\mathcal{M}^{2n+1}}{2n+1}
+\eta_{_A}\Big(1+\mathcal{M}^{-2} \Big)^{-3/2}
\right]^2\\
&=\frac{4 T C_v}{c^2}\mathcal{M}^3
\left\{
\Big(1+\mathcal{M}^2 \Big)^{3/2}
\left[
\sum_{n=0}^{\infty} \frac{\mathcal{M}^{2n}}{2n+3}
+\eta_{_A}\Big(1+\mathcal{M}^2 \Big)^{-3/2}
\right]^2\right\}.
\end{split}
\ee
Given that $C_v=2.01 \cdot 10^4\ \frac{J}{kg\ ^oK}$, we rewrite (\ref{3.31}) as
\be
\label{3.32}
\frac{\eta_{_A} \eta_{_{\Gamma}} \eta_{_G}^2}{T_6}
\gtrsim  9 \cdot 10^{-7} \mathcal{M}^3 \mathfrak{F}\big(\mathcal{M},\eta_{_A}\big),
\ee
where
\be
\label{3.33}
\mathfrak{F}\big(\mathcal{M},\eta_{_A}\big)=
\Big(1+\mathcal{M}^2 \Big)^{3/2}
\left[
\sum_{n=0}^{\infty} \frac{\mathcal{M}^{2n}}{2n+3}
+\eta_{_A}\Big(1+\mathcal{M}^2 \Big)^{-3/2}
\right]^2.
\ee

The Mach number for which a MBH settles into a stable intrastellar orbit is such that the net force acting on MBH is $0$.  It can be obtained by solving an equation derived from (\ref{3.32}):
\be
\label{3.34}
\mathcal{N}=\frac{\eta_{_A} \eta_{_{\Gamma}} \eta_{_G}^2}{T_6}
=  9 \cdot 10^{-7} \mathcal{M}^3 \mathfrak{F}\big(\mathcal{M},\eta_{_A}\big),
\ee
Even though (\ref{3.34}) does not have an explicit solution, it can be solved by a recursive relation
\be
\label{3.35}
\mathcal{M}_0 \approx
\left[
\frac{1.1 \cdot 10^6\ \eta_{_A} \eta_{_{\Gamma}} \eta_{_G}^2}
{\mathfrak{F}\big(\mathcal{M}_0,\eta_{_A}\big) T_6}
\right]^{1/3}=\left[
\frac{1.1 \cdot 10^6}
{\mathfrak{F}\big(\mathcal{M}_0,\eta_{_A}\big) \mathcal{N}}
\right]^{1/3}.
\ee
As the above recursive relation is applied to the initial guess for $\mathcal{M}_0$ several times, the solution of (\ref{3.34}) is obtained with a high degree of accuracy.

We have calculated the approximate values of $\mathfrak{F}\big(\mathcal{M},\eta_{_A}\big)$.  They are tabulated below:
\begin{center}
  \begin{tabular}{|l|||l|l|l|l|l|l|l|l|l|l|l|l|l|l|l|}
  \hline
  $\eta_{_A}\diagdown \mathcal{M}$&0&.1&.2&.3&.4&.5&.6&.7&.8\\
    \hline
    0.0 & 0.11 & 0.11 & 0.12 & 0.14 & 0.17 & 0.22 & 0.29 & 0.43 & 0.71 \\
    0.1 & 0.19 & 0.19 & 0.20 & 0.22 & 0.25 & 0.30 & 0.39 & 0.54 & 0.83 \\
    0.2 & 0.28 & 0.29 & 0.30 & 0.32 & 0.35 & 0.40 & 0.49 & 0.65 & 0.96 \\
    \hline
  \end{tabular}
  \captionof{table}{Values of $\mathfrak{F}\big(\mathcal{M},\eta_{_A}\big)$ \label{T01}}
\end{center}

Given that all efficiencies are nonzero, we conclude that any MBH within stellar material will accelerate until it obtains Mach number $\mathcal{M}_0$ at which it will fall into a stable intrastellar orbit.  The Mach number $\mathcal{M}_0$ can be calculated from (\ref{3.35}) once we know gas redistribution, accretion and radiative efficiencies.  Some estimates for these efficiencies are presented in the following Section.

\section{Estimation of gas redistribution, accretion and radiative efficiencies}
\subsection{The value of $\eta_{_G}$}
In Appendix A  we perform some steps in calculating $\eta_{_G}$ for different MBH velocities and list the unsolved problems.  If the heated gas trail generated by the PBH is at least twice hotter then unperturbed stellar medium, then
%In Appendix A, we prove that
\be
\label{4.01}
\begin{split}
\eta_{_G}>0.3 \ \text{ for MBH travelling at subsonic speeds with $\mathcal{M}<0.8$.}
\end{split}
\ee
%In our opinion based on numerical calculations, $\eta_{_G}>0.5$ for subsonic regime, but proving this more rigorous bound would require an extensive calculation.
If the gas is heated only slightly, then $\eta_{_G}$ would be much smaller.

\subsection{The value of $\eta_{_A}$}

From (\ref{3.15}), we estimate $\eta_{_A}$ as
\be
\label{4.02}
\begin{split}
\eta_{_A} \approx \left(\frac{v_0^2+v_s^2}{v_0^2+v_{sr}^2}\right)^{3/2} \frac{\rho_r}{\rho},
\end{split}
\ee
where $v_{sr}$ is the sound speed at the accretion radius and $\rho_r$ is the density at the accretion radius.  Given that gas density is proportional to its pressure divided by temperature, we obtain
\be
\label{4.03}
\begin{split}
\eta_{_A} &\approx
\left(\frac{v_0^2+v_s^2}{v_0^2+v_{sr}^2}\right)^{3/2} \frac{\rho_r}{\rho}
=\left(\frac{v_0^2+v_s^2}{v_0^2+v_{sr}^2}\right)^{3/2}
\frac{\mathbf{P}_r\ T}{\mathbf{P}\ T_r}
=\left(\frac{\big(v_0/v_s\big)^2+1}{\big(v_0/v_s\big)^2+\big(v_{sr}/v_s\big)^2}\right)^{3/2}
\frac{\mathbf{P}_r\ T}{\mathbf{P}\ T_r}\\
&=\left(\frac{\mathcal{M}^2+1}{\mathcal{M}^2+\big(v_{sr}/v_s\big)^2}\right)^{3/2}
\frac{\mathbf{P}_r\ T}{\mathbf{P}\ T_r},
\end{split}
\ee
where $\mathcal{M}$ is the Mach number for the MBH moving through stellar material.  Notice that notation $M$ is taken by the MBH mass.
Temperature of undisturbed medium is denoted by $T$, while temperature at Bondi radius is denoted by $T_r$.  Pressure of undisturbed medium is denoted by $\mathbf{P}$, while pressure at Bondi radius is denoted by $\mathbf{P}_r$.  Notice, that notation $P$ is already taken by power.  The sound velocity within gas is proportional to the square root of temperature.  Thus,
\be
\label{4.04}
\left(\frac{v_{sr}}{v_s}\right)^2=\frac{T_r}{T}.
\ee
Substituting (\ref{4.04}) into (\ref{4.03}), we obtain the approximation
\be
\label{4.05}
\eta_{_A} \approx
\left(\frac{\mathcal{M}^2+1}{\mathcal{M}^2+T_r/T}\right)^{3/2}
\frac{\mathbf{P}_r\ T}{\mathbf{P}\ T_r}.
\ee

The pressure within the immediate vicinity of MBH should be approximated by the sum of gas pressure and dynamic pressure:
\be
\label{4.06}
\mathbf{P}_r=\mathbf{P}+\frac{\rho v_0^2}{2}=\mathbf{P}+ \rho v_s^2\  \frac{\mathcal{M}^2}{2}.
\ee
This approximation is valid only while
\be
\label{4.07}
\frac{\mathbf{P}_r\ T}{\mathbf{P}\ T_r}<1,
\ee
since the density of the heated gas at Bondi radius cannot exceed the density of unperturbed gas.  For monatomic ideal gas, the pressure can be expressed in terms of the sound speed \cite[p.683]{soundv}:
\be
\label{4.08}
\mathbf{P}=\frac{\rho v_s^2}{\gamma}=\frac{3}{5} \ \ \rho v_s^2.
\ee
Substituting (\ref{4.08}) into (\ref{4.06}), we obtain the pressure ratio
\be
\label{4.09}
\frac{\mathbf{P}_r}{\mathbf{P}}=
\frac{\frac{3}{5} \ \ \rho v_s^2+\rho v_s^2\  \frac{\mathcal{M}^2}{2}}
     {\frac{3}{5} \ \ \rho v_s^2}=
     1+\frac{5}{6}\mathcal{M}^2.
\ee
Substituting (\ref{4.09}) and (\ref{4.07}) into (\ref{4.05}), we obtain
\be
\label{4.10}
\eta_{_A} \approx
\left(\frac{\mathcal{M}^2+1}{\mathcal{M}^2+T_r/T}\right)^{3/2}
\min \left\{1, \left(1+\frac{5}{6}\ \mathcal{M}^2\right)\frac{T}{T_r} \right\}.
\ee

As we see, $\eta_{_A}$ is a rapidly increasing function of the Mach number and a rapidly decreasing function of $T_r/T$.  For subsonic MBH and for all cases where
$T_r/T \gg \mathcal{M}^2$, (\ref{4.10}) can be approximated as
\be
\label{4.11}
\begin{split}
\eta_{_A} &\approx
\left(1+\frac{5}{6}\ \mathcal{M}^2\right)
\Big(\mathcal{M}^2+1\Big)^{3/2}
\left(\frac{T}{T_r}\right)^{5/2}.
\end{split}
\ee

The calculation of $T_r$ remains an unsolved problem.  As we see in the next Subsection, the radiative efficiency of spherically accreting black hole has not been calculated.  Different scientists working on the problem obtained different results.

%\begin{center}
%  {\rr \Huge Out of place: \normalsize}
%\end{center}
%\begin{quote}
%Substitute $\eta_{_{\Gamma}}=0.008$, $\eta_{_G}=0.3$, $\eta_{_A} = 3 \cdot 10^{-5}$ and
%$T_6=16$ into (\ref{3.35}).  We find that the minimum Mach number for a MBH on an intrastellar orbit within a Sun-like star is $\mathcal{M}_0=0.24$.

%We conclude this subsection with the following statement.  Any MBH trapped within a star experiences acceleration if it travels at Mach number $\mathcal{M}< \mathcal{M}_0$.  In almost all cases, such MBH would settle into an intrastellar orbit with Mach numbers $\mathcal{M}_0 \in (0.24, 0.8)$.  For a sun-like star, an orbit with $\mathcal{M}_0=0.24$ would have a radius of $0.07$ solar radii and an orbit with $\mathcal{M}_0=0.8$ would have a radius of $0.12$ solar radii (calculated based on data in \cite{solar}).
%\end{quote}

\subsection{The value of $\eta_{_{\Gamma}}$}
Some of the energy is radiated from a spherically accreting MBH in the form of photons.  The power radiated as photons is given as $\eta_{_{\gamma}} \dot{M} c^2$.  Some of the energy is radiated from a spherically accreting MBH in the form of protons and neutrons.  The power radiated as baryons is given as
$\eta_{_p}\dot{M} c^2 $, since protons are more numerous then neutrons.  The overall radiative efficiency of an MBH is
\be
\label{4.12}
\eta_{_{\Gamma}}=\eta_{_{\gamma}}+\eta_{_p}.
\ee

\subsubsection{Gamma radiation from spherically accreting MBH}

The radiative efficiency $\eta_{_{\gamma}}$ is dependent on accretion rate per unit MBH mass.  This rate is expressed in the units of Eddington accretion rate
\cite[p.51]{p25_55}.  It is
\be
\label{4.13}
\mathfrak{A}
=\frac{\Big(1.43 \cdot 10^{16}\ s \Big)\ \dot{M}}{M}
=\frac{\dot{M}}{\left( 70\ \frac{kg}{s} \right) M_{18}}.
\ee

Below we summarize some works calculating $\eta_{_{\gamma}}$ for spherical accretion.
Spherical accretion on black holes have been studied theoretically with different theories producing different values of radiative efficiency $\big(\eta_{_{\gamma}}\big)$ \cite[p.25-55]{p25_55}.  Radiative efficiencies ranging from $10^{-10}$ to over $.1$ have been obtained for different parameters.  Magnetic field greatly increases $\eta_{_{\gamma}}$ \cite[p.34-35]{p25_55}.  For
$10^{-4} \le \mathfrak{A} \le 1$, radiative efficiency can be as high as $0.1$ if the flow is turbulent \cite[p.35]{p25_55}.

Detailed calculations of spherical accretion are presented in \cite{Ind2018}.  For a black hole of $2 \cdot 10^{38}\ kg$, radiative efficiency starts growing almost from zero at $\mathfrak{A}=.02$ and reaches $\eta_{_{\gamma}}=.19$ for $\mathfrak{A}=1.2$.  For a black hole of $2 \cdot 10^{31}\ kg$, radiative efficiency starts growing almost from zero at $\mathfrak{A}=.5$ and reaches $\eta_{_{\gamma}}=.15$ for $\mathfrak{A}=.12$.  MBH were not considered.

A model which considers separate ion and electron temperatures within accreting gas is given in \cite{twotemp}.  Black hole masses between $2 \cdot 10^{31}\ kg$ and $2 \cdot 10^{38}\ kg$ are considered.  Accretion rates between $\mathfrak{A}=7 \cdot 10^{-3}$ and $\mathfrak{A}=2$ are considered.  In all cases, the efficiency stays within\\
 $\eta_{_{\gamma}} \in \Big[4.8 \cdot 10^{-3}, 7 \cdot 10^{-3} \Big]$.  Notice, that none of the aforementioned studies have worked with MBH with mass close to $10^{18}\ kg$.  Even though we can extrapolate the results from heavier black holes, extra calculations are needed.

Energy efficiency of spherical accretion on black holes still has a lot of uncertainty.  By 2013, different researchers found different $\eta_{_{\gamma}}$ for
$\mathfrak{A} \in (1,300)$ with results ranging from $10^{-6}$ to $10^{-2}$ \cite[p.10]{menu}.

\subsubsection{Proton and neutron radiation from spherically accreting MBH}

As the gas is undergoing spherical accretion toward MBH, it undergoes significant compression and adiabatic heating.  The resulting high temperature allows some protons and neutrons to escape from the accreting gas.  Below we perform a very rudimentary estimate of $\eta_{_p}$.  A precise calculation of $\eta_{_p}$ would consist of an extensive Monte Carlo simulations of proton motion and collisions.

The average temperature of electrons falls far short of the average temperature of protons.  The average temperature of protons can be approximated by (\cite[p.17]{Ind2018}, \cite[p.323]{twotemp}):
\be
\label{4.14}
T \big( y r_s \big)= \frac{T_s}{y},
\ee
where $r_s$ is the Schwarzschild radius and $T_s \approx 10^{12}\ ^oK$.  When the distance from MBH is corresponding to $y \in (25,50)$ Schwarzschild radii and the gas temperature is 20 to 40 billion Kelvin, the nuclei split into protons and neutrons.

At this point we calculate the depth of potential well in which protons and neutrons appear at a distance $\big( y r_s \big)$ from MBH center.  We take the non-relativistic approximation valid for $y \ge 2$.
\be
\label{4.15}
E_p \big( y r_s \big)
=-\frac{m_p\ M G}{y r_s}
=-\frac{m_p\ M G}{y \frac{2 M G}{c^2}}
=-\frac{m_p\ c^2}{2 y},
\ee
where $m_p$ is the proton mass.  Below we express (\ref{4.15}) in terms of Boltzmann constant\\ $k=1.381 \cdot 10^{-23}\ J/K$:
\be
\label{4.16}
E_p \big( y r_s \big)
=-\frac{m_p\ c^2}{2 y}
=-\frac{k}{y} \cdot \frac{m_p\ c^2}{2 k}
=-5.44\cdot 10^{12}\ ^oK\ \frac{k}{y}
\approx -\frac{5.44\ k\ T_s}{y}.
\ee

Assume protons and neutrons have Maxwell energy distribution measured with respect to the accreting gas,
\be
\label{4.17}
f(E)=\frac{2}{\sqrt{\pi}\ kT} \sqrt{\frac{E}{kT}}\ \exp \left( -\frac{E}{kT} \right),
\ee
At any point $y$ along the gas accretion into  MBH, a small fraction of protons and neutrons has enough kinetic energy to overcome the potential well given in (\ref{4.16}).  That fraction of protons capable of escaping is
\be
\label{4.18}
\mathcal{F}_e= \int_{5.44}^{\infty} f(E) dE =0.012.
\ee
The fraction of protons and neutrons capable of escaping from distance $\big( y r_s \big)$ carries an excess of kinetic energy given by
\be
\label{4.19}
\mathcal{F}_e
= \frac{k T_s}{y}\int_{5.44}^{\infty}(E-5.44) f(E) dE
=0.009\ \frac{k T_s}{y}.
\ee
The above energy is measured with respect to every proton or neutron, wether it escapes or not.

Let $\eta_{_p}^*>\eta_{_p}$ be the quotient of the energy of the protons and neutrons spit out of accreting material to the total rest energy of protons and neutrons.  A large fraction of these protons and neutrons will experience collisions with accreting protons and neutrons and thus they will return to the MBH.  We can estimate $\eta_{_p}^*$ as
\be
\label{4.20}
\frac{d \eta_{_p}^*}{d ln y}
=\frac{\mathcal{F}_e(y)}{m_p c^2}
=\frac{0.009\ k T_s}{m_p c^2} \frac{1}{y}
\approx \frac{8 \cdot 10^{-4}}{y}.
\ee
Integrating (\ref{4.20}) for $y>2$, we obtain
\be
\label{4.21}
\eta_{_p}^*
\approx
\int_{y=2}^{\infty} \frac{8 \cdot 10^{-4}\ dy}{y}
=4 \cdot 10^{-4}.
\ee

The value of $\eta_{_p}$ depends on the fraction of the protons and neutrons which are slowed down by accreting gas and return to MBH.  From the simplified model it would seem that $\eta_{_p} \ll \eta_{_{\gamma}}$.  We cannot rule out a thorough simulation giving a considerably higher value of $\eta_{_p}$.

\section{Possible modes of interaction of MBH with a star}
We consider the behavior of a Primordial Black Hole (PBH) which falls into an orbit intersecting a star.  Notice that the set of Primordial Black Holes described in this work is a subset of MBH.  Every PBH considered here is microscopic, while MBH considered here may or may not be primordial.

According to the classical theory \cite[p.8]{tidal}, a PBH is kicked into a highly elliptic orbit which contains an intrastellar segment.  As PBH orbits the stellar center, it decelerates with every intrastellar passage and eventually settles at the stellar center.

The first mode of PBH iteraction with a star is \textbf{PBH ejection}.  As we have shown in this work, any PBH or MBH passing through a star from a high-apogee orbit moves through stellar material at supersonic speed and experiences deceleration.  Nevertheless, a PBH or MBH orbiting a star on a high-apogee star-intersecting orbit is likely to be ejected from such orbit due to gravitational effect of the star's planetary system.  Reasoning follows.

According to our calculation in Appendix B, a MBH experiences an energy loss of
\be
\label{5.01}
\triangle E_{_{\text{pass}}} =\Big( 2.0 \cdot 10^{19}\ J\Big) M_{18}^2,
\ee
over each passage through a sun-like star.  The energy needed to drop the apogee of an elliptic MBH orbit around a sun-like star to 1 Astronomical Unit is
\be
\label{5.02}
\triangle E_{_{\text{orbit}}}= \frac{G M_{_\text{Sun}} M_{_{\text{MBH}}}}{1\ AU}=
\Big(8.9 \cdot 10^{26}\ J \Big) M_{18}.
\ee
Dividing (\ref{5.02}) by (\ref{5.01}), we obtain the number of times a MBH has to pass through a star in order for it's orbit apogee to descend to 1 AU:
\be
\label{5.03}
N=\frac{\triangle E_{_{\text{orbit}}}}{\triangle E_{_{\text{pass}}}}
\approx 4.5 \cdot 10^7 \big( M_{18} \big)^{-1}.
\ee
During this number of passes, the gravity of satellites of a star almost certainly throws a MBH off the orbit, but further calculations and simulations are needed to prove this point.

The second mode of MBH iteraction with a star is MBH settling into an intrastellar orbit.  That can happen in a rare event of stellar capture of PBH.  That can also happen in an event of MBH being produced within a star by coalescence of dark matter \cite{Type1a,wimp1}.

An MBH settling on an interstellar orbit may either experience small change in mass or consume the host star.  In order to determine which way an intrastellar MBH evolves,
we calculate its growth rate.  According to (\ref{3.13}), a stationary MBH will consume matter at the rate of
\be
\label{5.04}
\dot{M} =\frac{4 \pi (MG)^2 \rho_r}{v_{sr}^3},
\ee
where $\rho_r$ is the density at Bondi radius and and $v_{sr}$ is the sound speed an Bondi radius.  Given that the sound speed is proportional to the square root of temperature, and the density is inversely proportional to temperature, we obtain
\be
\label{5.05}
\dot{M} =\frac{4 \pi (MG)^2 \rho}{v_s^3}\ \left(\frac{T}{T_r}\right)^{2.5},
\ee
where $\rho$, $v_s$, and $T$ are the density, sound speed, and temperature at the stellar center, while $T_r$ is the temperature at Bondi radius.  In the Solar center, the density is
$1.5 \cdot 10^5\ \frac{kg}{m^3}$ and the sound speed is $5.1 \cdot 10^5\ \frac{m}{s}$ \cite[p.378]{solar}.  Substituting the above into (\ref{5.05}), we obtain the accretion rate
\be
\label{5.06}
\dot{M}
=6.3 \cdot 10^4\ \frac{kg}{s}\ \left(\frac{T}{T_r}\right)^{2.5}\ M_{18}^2
=\frac{2 \cdot 10^{18}\ kg}{\text{Million years}}\ \left(\frac{T}{T_r}\right)^{2.5}\ M_{18}^2.
\ee
Dividing both sides of the above equation by mass, we obtain
\be
\label{5.07}
\frac{d}{dt} \Big( \ln M_{18} \Big)=
\frac{\dot{M}_{18}}{M_{18}}=\frac{\dot{M}}{M}
=\frac{2 M_{18}}{\text{Million years}}\ \left(\frac{T}{T_r}\right)^{2.5},
\ee
where $T_r$ is the gas temperature at the Bondi radius.  If the MBH has low mass and high $T_r$, then the MBH may not grow by any appreciable amount during the star's lifetime.  Determination of exact dependence of $T_r$ on the PBH mass and stellar material density and temperature is beyond the scope of this work.

In cases where an MBH does consume the host star, it starts with slow growth.  As the MBH gains mass with $M_{18}>100$, all emitted radiation will be absorbed by the accreting gas and $T \approx T_r$.  Then the star will be consumed by MBH over several thousand years.

Growth of intrastellar black holes have been considered by previous researchers \cite{Tinyak}.  As a black hole consumes a star, it obtains the star's angular momentum and becomes a rapidly rotating black hole.  As a rotating black hole absorbs matter, it radiates two jets along its axis \cite{Jets}.  Final stages of stellar consumption by MBH may be responsible for long $\gamma$-ray pulses \cite{Gamma1}.

As far as we are aware, the possibility of MBH settling in an interstellar orbit has never been discussed before.  Possibly one or more intrasolar MBH will be discovered by \textbf{helioseismology} -- the science of studying solar structure by absorbing the vibrations of Sun's photosphere.

Interesting phenomena are likely to occur when stars with intrastellar MBH become white dwarves.  The MBH which have been orbiting within the star become extrastellar satellites with very small orbits.  Eventually, these MBH fall to the surface of the white dwarf due to tidal forces.  MBH falling into white dwarves may trigger Type 1a Supernova explosions \cite{Type1a}.

\section{Conclusion and Remaining Problems}

In this work we have demonstrated that MBH passing through stellar material will experience acceleration rather than deceleration as long as \be
\label{6.01}
\mathcal{N}=\frac{\eta_{_A} \eta_{_{\Gamma}} \eta_{_G}^2}{T_6}
\gtrsim
\left\{
\begin{split}
&4 \cdot 10^{-4}
\hskip3.35cm \text{for supersonic MBH speed}
\\
&9 \cdot 10^{-7} \mathcal{M}^3 \mathfrak{F}\big(\mathcal{M},\eta_{_A}\big)
\qquad
\text{for subsonic MBH speed}
\end{split}
\right|,
\ee
where $\mathcal{M}$ is the Mach number and
$\mathfrak{F}\big(\mathcal{M},\eta_{_A}\big) \in (0.11,1)$ is given in (\ref{3.33}).
$T_6$ is the temperature of the stellar material in millions Kelvin.
The
\textbf{gas redistribution efficiency} $\eta_{_G}$,
\textbf{radiative efficiency} $\eta_{_{\Gamma}}$, and
\textbf{radiative efficiency} $\eta_{_{\Gamma}}$
are defined in Introduction and Section 3.

MBH in stellar material experiences deceleration at supersonic speed.  MBH moving at subsonic speed settles into a stable intrastellar orbit with Mach number determined from (\ref{3.35}).

If Microscopic Black Holes exist in the Universe, many of them may have settled {\rr into} intrastellar orbits within the interiors of many stars including Sun.  First, there may be Primordial Black holes captured by stars.  As discussed in Section 5, stellar capture of PBH may be {\rr a rare phenomenon} due to gravitational effect of stellar planetary systems.  Nevertheless, such captures may occur.  Calculating the probability for stellar capture of free-floating MBH for different types of stars with different types of planetary systems is an open problem itself.  Second, Microscopic Black Holes may form within stellar interiors.  If all or part of Dark Matter consists of Weakly Interacting Massive Particles (WIMPs), then it is likely that they coalesce into MBHs within stellar interiors \cite{Type1a,wimp1}.  All MBHs within stellar interiors settle within intrastellar orbits.

MBH on intrastellar orbits may produce several physically observable effects.
First, these MBHs may trigger Type 1a supernovas \cite{Type1a}.  The minimal mass of MBHs capable of igniting Type 1a supernovas remains to be calculated.  Second, one or more MBHs may be orbiting within the Sun.  Sonic waves produced by these MBHs may be detectable using \textbf{helioseismology} -- the studying {\rr of the} solar structure by {\rr observing} the vibrations of Sun's photosphere.

Sound attenuation coefficient in gas grows rapidly with increasing frequency of the sound.
Only the sound of very low frequency can travel extended distances in any gas.
Most of us have observed that the sound of {\rr a} distant thunder has very low frequency. This is due to the fact that {\rr high} frequency components of thunder are attenuated on their passage through the atmosphere.  Infrasound can travel for hundreds of kilometers.  In order for sound to travel from Solar interior to the surface, it's frequency must be a few millihertz or lower \cite{HSeism01}.  Such waves have lengths of up to 10,000 $km$ in Solar atmosphere \cite[p.67]{HSeism02}.  Based on the data in \cite[p. 378]{solar} {\rr on the Solar material density}, the rotational period of an MBH within the Sun is at least 800 $s$.  Thus, the acoustic waves generated by MBH would be detectable.

In Subsection 4.3, we have mentioned that the \textbf{radiative efficiencies} of spherically accreting black holes are unknown.  Recall, that \textbf{radiative efficiency} of an accreting black hole is the ratio of the total power radiated by MBH to the energy $\big( M c^2 \big)$ of the mass falling into the black hole.  Most advanced theories give results, which  vary by several orders of magnitude.  Values ranging from $10^{-10}$ to 0.1 have been obtained so far.  We do not know, which theory is correct.  The only way to resolve the issue is by experimental investigation.  If one or more MBH orbiting within the Sun is detected, then true values of radiative efficiencies will be obtained from observation.  The knowledge of radiative efficiencies of MBH under different conditions may enable the researchers to prove the correct theory describing accretion and to reject the theories giving wrong answers.

This work and the hypothesis described therein are purely theoretical.  Nevertheless, if the hypothesis about MBH orbiting within Solar material is observationally validated, then we may obtain additional knowledge about interaction of particles at very high energies.  Even if extra knowledge in areas of physics which are considered purely theoretical at this point bring no immediate technological progress, such knowledge may bring technological advances in the coming decades.  {\rr In particular,} this knowledge may be especially useful in Colonization of Solar System, {\rr which is} the subject {\rr of interest for a broad scientific community}.

In Appendix A,  we perform some steps in calculating $\eta_{_G}$ and list some unsolved problems.
Exact calculation of $\eta_{_G}$ is a remaining problem.  It would involve extensive theoretical work and simulations using gas dynamics and radiation-matter interaction.

Previous results for radiative efficiency $\eta_{_{\Gamma}}$ are discussed in Section 4.3.  It is possible, that the correct result will be determined only when intrasolar MBH are detected and observational data is presented. Accretion efficiency $\eta_{_A}$ for both subsonic and supersonic MBH is given by (\ref{4.10}) in terms of $T_r$ -- the temperature at the Bondi radius.  The calculation of $T_r$ is a remaining problem.  Exact calculation of $T_r$, and $\eta_{_A}$ would involve extensive theoretical work and simulations using gas dynamics, radiation energy transport, and magnetohydrodynamics.  Possibly these problems will also remain unsolved until observational data become available. {\rr It is desirable to obtain experimental results to verify the hypothesis formulated in this work.}

\newpage
\appendix
\section{Estimation of bounds on $\eta_{_G}$}
\subsection{Minimum value of $\eta_{_G}$ for MBH moving at subsonic speed}
We assume strictly subsonic regime with $\mathscr{M} \le 0.8$.  In a diagram below, we describe the MBH passing through stellar material.
\begin{center}
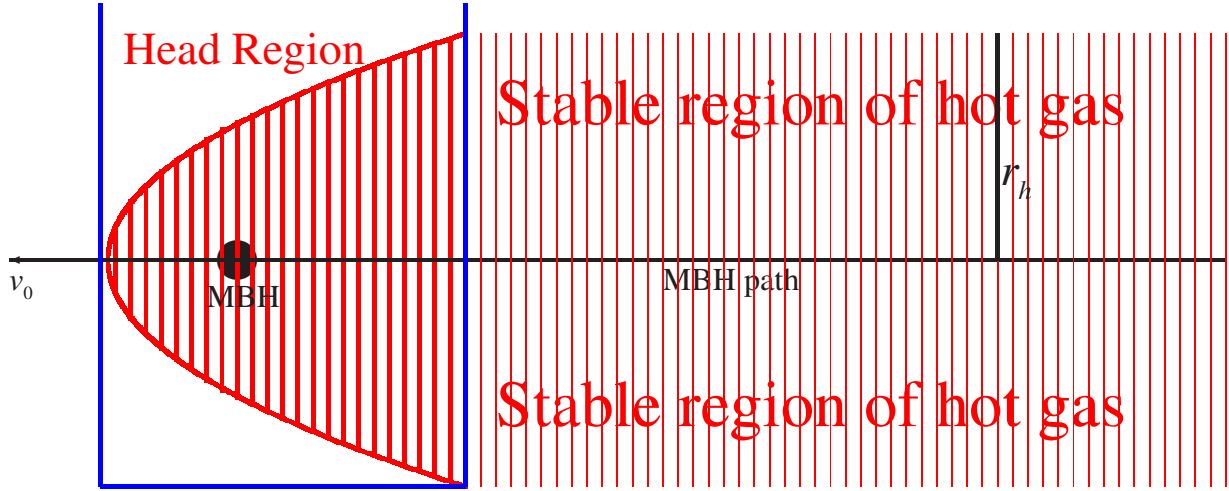

\setlength{\unitlength}{2mm}
\begin{picture}(90, 30)
\linethickness{0.5mm}
\put(15,15){\circle*{5}}
{\rr
\qbezier(30,0)(-17,15)(30,30)
\multiput(29,29.65)(-1,-0.35){9}{\line(0,-1){15}}
\multiput(29,0.35)(-1,0.35){9}{\line(0,1){15}}
\multiput(20,26.5)(-1,-0.45){6}{\line(0,-1){12}}
\multiput(20,3.5)(-1,0.45){6}{\line(0,1){12}}
\multiput(14,23.8)(-1,-.7){4}{\line(0,-1){9}}
\multiput(14,6.2)(-1,.7){4}{\line(0,1){9}}
\multiput(10,21.0)(-1,-1){3}{\line(0,-1){6}}
\multiput(10,9)(-1,1){3}{\line(0,1){6}}
\put(7,12.8){\line(0,1){4.5}}
}
\linethickness{0.3mm}
\put(10,15){\vector(-1,0){10}}
\put(13,11.9){MBH}
\put(0,13){$v_{_0}$}
\put(0,15){\line(1,0){80}}
\put(43,13){MBH path}
%\put(15,15){\line(0,1){9}}
%\put(15.3,20){$r_{_{hs}}$}
\linethickness{0.1mm}
\multiput(31,0)(1,0){50}{\rr{\line(0,1){30}}}
\put(32,24){\Huge \rr{Stable region of hot gas}}
\put(32,4){\Huge \rr{Stable region of hot gas}}
\linethickness{.5mm}
%\put(15,24){\rr \line(5,2){15}}
%\put(15,6){\rr \line(5,-2){15}}
\put(65,15){\line(0,1){15}}
\put(65.2,20){\Large $r_{_h}$}
%\put(15,15){\color{blue}{\line(0,-1){15}}}
\put(6,0){\color{blue}{\line(0,1){32}}}
%\put(6,6){\color{blue}{\line(1,0){9}}}
\put(30,0){\color{blue}{\line(0,1){32}}}
\put(6,0){\color{blue}{\line(1,0){24}}}
\put(7.5,28){\Large \rr{Head Region}}
\end{picture}
\captionof{figure}{Heat wave caused by MBH moving at a subsonic speed \label{F02}}
\end{center}

The heated stellar material produced by MBH moving at subsonic speed consists of two regions.  The first is the parabolic head region of hot gas surrounding the MBH.  The second is the hot gas trail.  The second region is denoted $\mathcal{R}_{hg}$.

An important issue is the location of stellar material mass displaced by the heat wave.  For an MBH travelling at supersonic speed, the displaced mass does not have time to move very far.  This greatly decreases the effect of rarefication and thus $r_1$ and $\eta_{_G}$.  In our case of MBH travelling at subsonic speed,
all of the extra mass of stellar material pushed out of the cylindrical region of hot gas is carried away by sonic density waves.
These waves are spherical shells.  Each shell's center is a point at which the wave originated.  Due to MBH's subsonic velocity,
the MBH is located inside of all the aforementioned spherical shell sound waves.
Thus, these waves exert no net gravitational pull on the MBH.

The force exerted on MBH comes from the fact that hot rarified gas both within the head region and within $\mathcal{R}_{hg}$ exerts lower gravitational pull on MBH than the dense stellar material in front of MBH.  Recall, that the total accelerative force is called $F_r$. Even though the head region contributes a small part of $F_r$, we ignore its contribution.

We calculate the force exerted by $\mathcal{R}_{hg}$.  The region $\mathcal{R}_{hg}$ can be approximated by a cylinder with radius $r_h$.  The cylinder starts at the distance at most $r_h$ from the MBH.  By taking the distance to be $r_h$, we are estimating minimal value of the force.  The region $\mathcal{R}_{hg}$ is represented in cylindrical coordinates with MBH at the origin.  The direction in which the MBH is travelling is $-\hat{z}$.  In cylindrical coordinates, $\mathcal{R}_{hg}$ is given by
\be
\label{A.01}
\left\{
\begin{split}
z & \in \big( r_h, \infty \big)\\
r & \in \big[ 0, r_h \big)
\end{split}
\right.
\ee

The temperature distribution of gas in $\mathcal{R}_{hg}$ is approximated by a uniform temperature $KT$, where $T$ is the temperature of surrounding material, and $K>1$ is a constant.  The pressure of gas in $\mathcal{R}_{hg}$ is approximately the same as that of surrounding stellar medium.  The density of the gas in $\mathcal{R}_{hg}$ is $\rho/K$, where $\rho$ is the density of surrounding stellar material.
The \textbf{effective negative density} $\rho_-$ of the material in $\mathcal{R}_{hg}$ is the difference of density of material in $\mathcal{R}_{hg}$ and the density of surrounding material:
\be
\label{A.02}
\rho_-=\frac{\rho}{K}-\rho=-\rho\ \frac{K-1}{K}.
\ee

The force acting on a MBH due to rarefied region $\mathcal{R}_{hg}$ is
\be
\label{A.03}
\textbf{F}
=\underset{\mathcal{R}_{hg}}{\iiint} \big(\rho_-\big) \textbf{g}(\mathbf{r})  dV
=-\rho\ \frac{K-1}{K} \underset{\mathcal{R}_{hg}}{\iiint}  \textbf{g}(\mathbf{r})  dV,
\ee
where
\be
\label{A.04}
\textbf{g}(\mathbf{r})=-M G \frac{\mathbf{r}}{\|\mathbf{r}\|^3}
\ee
is the acceleration due to MBH gravity at point $\mathbf{r}$.  Substituting (\ref{A.04}) into (\ref{A.03}), we obtain
\be
\label{A.05}
\textbf{F}
=-M G \rho\ \frac{K-1}{K}
\underset{\mathcal{R}_{hg}}{\iiint}
\frac{\mathbf{r}}{\|\mathbf{r}\|^3}  dV,
\ee

The gas displaced by the MBH passage in $-\hat{z}$ direction retains cylindrical symmetry.
This symmetry implies that the net force on the MBH will act only in $-\hat{z}$ direction.
Thus, (\ref{A.05}) can be further simplified to
\be
\label{A.06}
\begin{split}
F_r&=-\textbf{F} \cdot \hat{z}
=M G \rho\ \frac{K-1}{K}
\underset{\mathcal{R}_{hg}}{\iiint}
\frac{\mathbf{r}\cdot \hat{z}}{\|\mathbf{r}\|^3}  dV
=M G \rho\ \frac{K-1}{K}
\underset{\mathcal{R}_{hg}}{\iiint}
\frac{z}{\big(z^2+r^2\big)^{3/2}}\  dV\\
& \ge
M G \rho\ \frac{K-1}{K}
\int_0^{r_h} \int_{r_h}^{\infty}
\frac{\pi r z}{\big(z^2+r^2\big)^{3/2}} dz dr
=\pi M G \rho\ \frac{K-1}{K}
\int_0^{r_h}
\left[
-\frac{r}{\sqrt{z^2+r^2}}
\right]_{z=r_h}^{z=\infty} dr\\
&= \pi M G \rho\ \frac{K-1}{K}
\int_0^{r_h}
\frac{r}{\sqrt{r_h^2+r^2}} dr
=
\pi M G \rho\ \frac{K-1}{K}
\left[ \sqrt{r_h^2+r^2} \right]_{r=0}^{r_h} \\
&=\pi M G \rho\ r_h \left( \frac{K-1}{K} \big( \sqrt{2}-1 \big) \right).
\end{split}
\ee
As we have mentioned earlier, the real force is greater or equal to the one calculated by approximating $\mathcal{R}_{hg}$ by (\ref{A.01}).
Recall (\ref{3.07}):
\be
\label{A.07}
F_r=\frac{2}{3} \pi M G \rho r_1.
\ee
From (\ref{A.06}) and (\ref{A.07}), it follows that
\be
\label{A.08}
r_1 \ge  \frac{3}{2}\big( \sqrt{2}-1 \big)\ \frac{K-1}{K}\ r_h
\approx 0.62\ \frac{K-1}{K}\ r_h.
\ee

Below, $r_h$ is estimated in terms of $r_2$. The power needed to heat the gas trail is
\be
\label{A.09}
\begin{split}
P_{_T}&=\big(\text{Mass heated per unit of time} \big) \cdot \big(\text{Temperature}\big) \cdot C_p\\
&=v_0 \left(\pi r_h^2 \frac{\rho}{K} \right) \big((K-1)T\big)
\left(\frac{5}{3} C_v\right)
= \frac{5(K-1)}{3 K} \pi v_0 \rho T C_v r_h^2,
\end{split}
\ee
where $P_{_T}$ is the thermal power and $C_p$ is the heat capacity of the gas at constant pressure.  Recall that for monatomic gas, $C_p=\frac{5}{3} C_v$.  The total power radiated by the MBH must be greater, since some of it goes into the production of the sound waves.  Even though determining the  thermal efficiency of the sonic boom production by a MBH moving through the stellar material is beyond the scope of this work, it can be assumed that it does not exceed 20\% for subsonic MBH.  Hence, the total radiative power of MBH is
\be
\label{A.10}
P \le  \frac{2(K-1)}{K} \pi v_0 \rho T C_v r_h^2
\ee
Recall (\ref{3.08}):
\be
\label{A.11}
P=\pi v_0 \rho T C_v r_2^2.
\ee

Combining (\ref{A.10}) and (\ref{A.11}), we obtain
\be
\label{A.12}
\frac{2(K-1)}{K} \pi v_0 \rho T C_v r_h^2 \ge  \pi v_0 \rho T C_v r_2^2.
\ee
Hence,
\be
\label{A.13}
r_2 \le  r_h\sqrt{\frac{2(K-1)}{K}}.
\ee

Dividing (\ref{A.08}) by (\ref{A.13}), we obtain an estimate for the thermal efficiency for an MBH moving at subsonic speed
\be
\label{A.14}
\eta_{_G}=\frac{r_1}{r_2} \ge  .44\ \sqrt{\frac{K-1}{K}}.
\ee
Calculation of $K$ is beyond the scope of this work. Some considerations regarding the value of $K$ are presented in Appendix A.3.

\subsection{Maximum value of $\eta_{_G}$ for MBH moving at any speed}
The absolute maximum for $\eta_{_G}$ is obtained if the stable region of hot gas denoted $\mathcal{R}_{hg}$ starts right at MBH.  Such configuration is impossible, but it is useful for calculating the absolute maximum.  It is shown in Figure \ref{F03} below:
\begin{center}
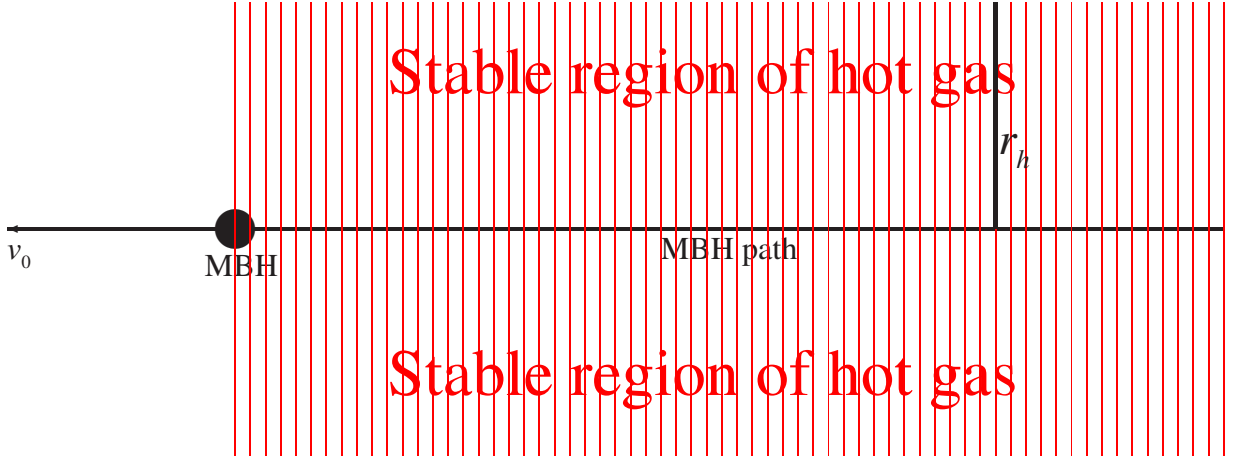

\setlength{\unitlength}{2mm}
\begin{picture}(90, 30)
\linethickness{0.5mm}
\put(15,15){\circle*{5}}

\linethickness{0.3mm}
\put(10,15){\vector(-1,0){10}}
\put(13,11.9){MBH}
\put(0,13){$v_{_0}$}
\put(0,15){\line(1,0){80}}
\put(43,13){MBH path}

\linethickness{0.1mm}
\multiput(15,0)(1,0){66}{\rr{\line(0,1){30}}}
\put(25,24){\Huge \rr{Stable region of hot gas}}
\put(25,4){\Huge \rr{Stable region of hot gas}}
\linethickness{.5mm}

\put(65,15){\line(0,1){15}}
\put(65.2,20){\Large $r_{_h}$}
\end{picture}
\captionof{figure}{Gas configuration for maximal $\eta_{_G}$ \label{F03}}
\end{center}
Recalling (\ref{A.06}), we evaluate the force
\be
\label{A.15}
\begin{split}
F_r&=-\textbf{F} \cdot \hat{z}
=M G \rho\ \frac{K-1}{K}
\underset{\mathcal{R}_{hg}}{\iiint}
\frac{\mathbf{r}\cdot \hat{z}}{\|\mathbf{r}\|^3}  dV
=M G \rho\ \frac{K-1}{K}
\underset{\mathcal{R}_{hg}}{\iiint}
\frac{z}{\big(z^2+r^2\big)^{3/2}}\  dV\\
& \le  M G \rho\ \frac{K-1}{K}
\int_0^{r_h} \int_{0}^{\infty}
\frac{\pi r z}{\big(z^2+r^2\big)^{3/2}} dz dr
=\pi M G \rho\ \frac{K-1}{K}
\int_0^{r_h}
\left[
-\frac{r}{\sqrt{z^2+r^2}}
\right]_{z=0}^{z=\infty} dr\\
&=\pi M G \rho\ \frac{K-1}{K}
\int_0^{r_h}
 dr
=
\pi M G \rho\ r_h \frac{K-1}{K}.
\end{split}
\ee
The real force is less or equal to the one calculated by approximating $\mathcal{R}_{hg}$ by Figure \ref{F03}.
From (\ref{A.15}) and (\ref{A.07}), it follows that
\be
\label{A.16}
r_1= 1.5\ \frac{K-1}{K}\ r_h.
\ee

Dividing (\ref{A.16}) by (\ref{A.13}), we obtain an upper limit for the thermal efficiency
\be
\label{A.17}
\eta_{_G}=\frac{r_1}{r_2} \le 1.06\ \sqrt{\frac{K-1}{K}}.
\ee
%It follows that $\eta_{_G}<1.06$.

We conjecture that, the real values of $\eta_{_G}$ are much lower, and they decrease with Mach Number.  We would estimate $\eta_{_G}$ as
\be
\label{A.18}
\begin{split}
\eta_{_G} \approx \frac{0.8}{M} \ \sqrt{\frac{K-1}{K}} \ \text{ for MBH travelling at Mach } \ge 2,\\
\end{split}
\ee
but until there are rigorously derived and proven results, it remains an open problem.

\subsection{Considerations regarding the value of $K$}
Recall, that the average temperature of the gas in the hot tail is $K T$, where $T$ is the temperature of the surrounding stellar material.  The exact calculation of $K T$ is beyond the scope of this work, and possibly beyond the scope of modern science.  Nevertheless, here are some general observations on $K$.

We introduce two new constants.  \textbf{Radiation radius}  $r_{_{\gamma}}$ is the average distance travelled by a photon or another energy-carrying particle from PBH before being absorbed by stellar material.  \textbf{Minimal hot tail radius} $r_{_{mh}}$ is the minimal radius the hot tale can have regardless of $K$.

If $r_{_{\gamma}} \ll r_{_{mh}}$, then gas close to MBH is heated to a great temperature.  This gas expands before it has time to diffuse its heat.  The expanded gas must remain hot in order to balance the outside pressure.  In that case, $K \gg 1$.  If $r_{_{\gamma}} \gg r_{_{mh}}$, then heat is dissipated over a large volume of gas.  That gas volume is much larger than the gas volume which can be significantly heated by MBH radiative power.   Thus, the heating has to be by a small margin.  Thus,
$0 < K-1 \ll 1$.

At this point we introduce some calculations for estimating $r_{_{\gamma}}$  and $r_{_{mh}}$.  The radiation radius is
\be
\label{A.19}
 r_{_{\gamma}}=\frac{\mathfrak{S}_{_{\gamma}}}{\rho_{_p}},
\ee
where $\mathfrak{S}_{_{\gamma}}$ is the planar density of material through which an energy carrying particle has to travel before being absorbed by stellar material.  Recall that absorbtion of any radiation in any medium is proportional to planar density.  The density $\rho_{_p}$ is an average density of the material over the path of the energy-carrying particle.  The value of $\mathfrak{S}_{_{\gamma}}$ is inversely proportional to average absorbtion cross-section of the energy-carrying particles:
\be
\label{A.20}
\mathfrak{S}_1
= \frac{1\ kg}{1000 N_A\ amu} \cdot \frac{1}{10^{-28}\ \sigma}
=\frac{16.6\ \frac{kg}{m^2}}{\sigma\ \text{in barn}},
\ee
where $\sigma$ is the absorbtion cross-section in barns.  Given that most interactions are scattering, effective absorbtion cross-section has to be calculated.

We calculate the attenuation coefficient of photons.  In Table \ref{T02} below, the first column is the photon energy.  The second column is the total photon absorbtion and scattering cross-section per amu  \cite[pp.41-42]{gamma}.  The third column is the mass attenuation coefficient.  It is calculated in (\ref{A.20}).  For low energy photons, all of the events are scattering.  For high energy photons, there is another algorithm for interaction with matter -- production of $\left\{ e^-, e^+ \right\}$ pairs.  The forth column denoted PPF is the pair production fraction of all events.
\begin{center}
\begin{tabular}{|c|c|c|c|}
  \hline
  $E(\gamma)$ & $\sigma$ in barn            & $\mathfrak{S}_{\gamma}$ & PPF\\
              & $\Big( 10^{-28}\ m^2 \Big)$ & in $kg/m^2$             & \\
  \hline
  $10\ keV$   & 0.55  &  30 & 0 \\
  $20\ keV$   & 0.53  &  31 & 0 \\
  $50\ keV$   & 0.48  &  35 & 0 \\
  $100\ keV$  & 0.42  &  40 & 0 \\
  $200\ keV$  & 0.35  &  47 & 0 \\
  $500\ keV$  & 0.25  &  66 & 0 \\
  $800\ keV$  & 0.20  &  83 & 0 \\
  $1.0\ MeV$  & 0.18  &  92 & 0 \\
  $1.5\ MeV$  & 0.146 & 114 & 0.0003 \\
  \hline
\end{tabular}
\begin{tabular}{|c|c|c|c|}
  \hline
  $E(\gamma)$ & $\sigma$ in barn            & $\mathfrak{S}_{\gamma}$ & PPF \\
              & $\Big( 10^{-28}\ m^2 \Big)$ & in $kg/m^2$             &\\
  \hline
    $2.0\ MeV$  & 0.124 & 134 & 0.0012 \\
    $3.0\ MeV$  & 0.099 & 170 & 0.0047 \\
    $5.0\ MeV$  & 0.073 & 230 & 0.016 \\
    $8.0\ MeV$  & 0.054 & 310 & 0.041 \\
    $10\ MeV$   & 0.048 & 350 & 0.062 \\
    $15\ MeV$   & 0.038 & 440 & 0.11 \\
    $20\ MeV$   & 0.032 & 520 & 0.16 \\
    $30\ MeV$   & 0.027 & 620 & 0.24 \\
    $50\ MeV$   & 0.023 & 720 & 0.40 \\
  \hline
\end{tabular}
\captionof{table}{Photon mass attenuation coefficient \label{T02}}
\end{center}

The minimal hot tail radius can be obtained from (\ref{A.09}):
\be
r_{_{mh}}=\sqrt{\frac{3 P_{_T}}{5 \pi v_0 \rho T C_v}},
\ee
where $P_{_T}$ is the part of MBH power used to produce heat rather than the sound wave.

\section{Estimation of a MBH kinetic energy loss on passage through a sun-like star}
This subsection consists of mostly numerical calculations.  In our calculations we use the force given in (\ref{3.01}), which is given in \cite[p.8]{tidal} and also derived in Appendix C.  We use $r_{_{\text{min}}}=0.1\ m$ and
$r_{_{\text{max}}}=5 \cdot 10^7\ m$.
Expressing (\ref{3.01}) in numerical terms we obtain:
    \be
    \label{B.01}
    F_t=\frac{4 \pi \big( M G \big)^2 \rho}{v_0^2}\
    \ln \left( \frac{r_{_{\text{max}}}}{r_{_{\text{min}}}} \right)
    =\big( 1.12 \cdot 10^9\ N\big)\ \frac{M_{18}^2 \ \rho_3}{v_6^2}.
    \ee

Below we tabulate several parameters for a MBH passing through a sun-like star.  We use the density data from Solar interior given in \cite{solar}.  The first column stands for the fraction of Solar radius denoted by $R_{\text{Sun}}$.  The second column is the density in $g/cm^3$.  The third column is an estimated speed of a MBH which arrives at the sun from a large distance.  The forth column is $F_t$ for $M_{18}=1$.
\begin{center}
  \begin{tabular}{|c|c|c|c|c|c|c|c|c|c|c|c|c|c|c|c|}
     \hline
      $R_{\text{Sun}}$   & $\rho_3$   & $v_6$ & $F_t/M_{18}^2$\\
     \hline
     0.0 & 146   & 1.39 & $85 \cdot 10^9\ N$\\
     0.1 &  82   & 1.33 & $51 \cdot 10^9\ N$\\
     0.2 &  35   & 1.19 & $28 \cdot 10^9\ N$\\
     0.3 & 12.3  & 1.06 & $12.3 \cdot 10^9\ N$\\
     0.4 & 4.0   & 0.96 & $4.9 \cdot 10^9\ N$\\
     0.5 & 1.35  & 0.87 & $2.0 \cdot 10^9\ N$\\
     0.6 & 0.49  & 0.80 & $0.86 \cdot 10^9\ N$\\
     0.7 & 0.185 & 0.74 & $0.38 \cdot 10^9\ N$\\
     0.8 & 0.077 & 0.69 & $0.18 \cdot 10^9\ N$\\
     \hline
   \end{tabular}
   \captionof{table}{Parameters for MBH passing through a Sun-like star}
\end{center}

The Solar radius is $R_{\odot}=6.96 \cdot 10^8\ m$.  Thus we estimate the energy loss of a MBH passing through the center of a Sun-like star:
    \be
    \label{B.02}
    \triangle E
    =\int_{-R_{\odot}}^{R_{\odot}} F_t\ dx
    =\Big( 2.0 \cdot 10^{19}\ J\Big) M_{18}^2.
    \ee

\section{Derivation of (\ref{3.01})}
In this section we show that a small black hole passing through gas at supersonic speed experiences a drag force given in (\ref{3.01}).
%      \be
%      \label{C.01}
%      F_t=-\frac{4 \pi (MG)^2 \rho}{v_0^2}
%      \ln \left(\frac{r_{\max}}{r_{\min}}\right),
%      \ee
%where $P_t$ is the decelerating power exerted by drag force, $\rho$ is the density of the surrounding medium, $r_{\max}$ is the radius of the star, and $r_{\min}$ is the radius at which accretion begins.
\begin{center}
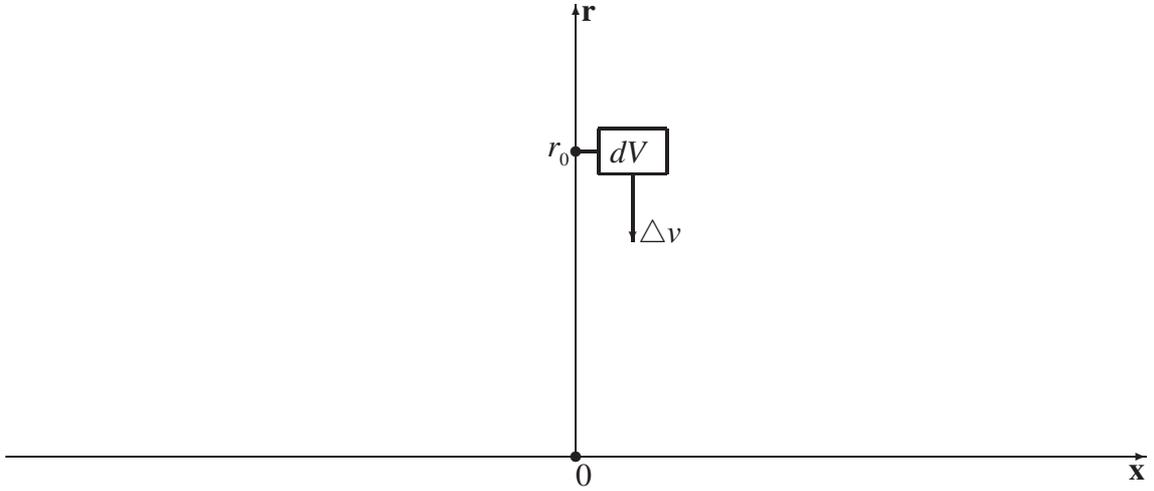

\setlength{\unitlength}{1.5mm}
\begin{picture}(100, 50)
    \put(0,5){\vector(1,0){100}}
    \put(50,5){\vector(0,1){40}}
    \put(50,5){\circle*{1}}
    \put(50,2.5){$0$}
    \put(98.5,3){$\mathbf{x}$}
    \put(50.5,43.5){$\mathbf{r}$}
    \linethickness{0.3mm}
    \put(52,30){\line(0,1){4}}
    \put(58,30){\line(0,1){4}}
    \put(52,30){\line(1,0){6}}
    \put(52,34){\line(1,0){6}}
    \put(53,31){$dV$}
    \put(50,32){\line(1,0){2}}
    \put(50,32){\circle*{1}}
    \put(47.5,31.5){$r_{_0}$}
    \put(55,30){\vector(0,-1){6}}
    \put(55.5,24){$\triangle v$}
\end{picture}
\captionof{figure}{Effects of black hole passage through gas}
\end{center}
\par The black hole passes along the $\mathbf{x}$ -- axis in $\hat{x}$ direction.  The black hole $x$ -- coordinate is the function of time:
    \be
    \label{C.01}
    x(t)=v_0 t.
    \ee
During the passage, it changes the speed of gas by providing a push into $-\hat{r}$ direction.  For any volume $dV$ of gas located at the distance $r$ from x-axis, the speed supplied by the gravitational push is
    \be
    \label{C.02}
    \begin{split}
    \triangle v&
    =\int_{-\infty}^{\infty} \Big( \text{Acceleration in $-\hat{r}$ direction at time $t$} \Big) dt
    =\int_{-\infty}^{\infty} MG \frac{r}{\left(r^2+x^2\right)^{1.5}}
    \left( \frac{dx}{v_0} \right)\\
    &=\frac{MG}{v_0} \int_{-\infty}^{\infty}
    \frac{r\ dx}{\left(r^2+x^2\right)^{1.5}}
    =\frac{MG}{v_0\ r} \int_{-\infty}^{\infty}
    \frac{d \big( x/r \big)}{\left(1+\big(x/r\big)^2\right)^{1.5}}
    =\frac{2MG}{v_0\ r}.
    \end{split}
    \ee

The energy supplied to the gas within the volume $dV$ is
    \be
    \label{C.03}
    \triangle E
    =\Big( \rho \triangle V \Big) \frac{\big(\triangle v\big)^2}{2}
    =\frac{2 \big(M G \big)^2}{v_0^2\ r^2}\ \rho \ \triangle V.
    \ee
The energy supplied to a cylindrical shell of length $dx$ and thickness $dr$ can be expressed by
    \be
    \label{C.04}
    \frac{dE}{dr\ dx}
    =\triangle E \cdot \big( \text{Area} \big)
    =\left( \frac{2 \big(M G \big)^2}{v_0^2\ r^2}\ \rho \right) \cdot
    \big( 2 \pi r \big)
    =\frac{4 \pi \big( M G \big)^2 \rho}{v_0^2\ r}.
    \ee
Integrating the above we obtain the resistive drag force:
    \be
    \label{C.05}
    F_t=\frac{dE}{dx}= \int_{-\infty}^{\infty} \frac{dE}{dr\ dx} dr
    =\int_{-\infty}^{\infty} \frac{4 \pi \big( M G \big)^2 \rho}{v_0^2\ r} dr
    =\frac{4 \pi \big( M G \big)^2 \rho}{v_0^2}\
    \ln \left( \frac{r_{_{\text{max}}}}{r_{_{\text{min}}}} \right).
    \ee
The power is
    \be
    \label{C.06}
    P=v_0\ \frac{dE}{dx}
    =\frac{4 \pi \big( M G \big)^2 \rho}{v_0}\
    \ln \left( \frac{r_{_{\text{max}}}}{r_{_{\text{min}}}} \right).
    \ee

Notice, that the above result is valid if and only if the flow is supersonic.  For transonic and subsonic flows, the gas is stopped by pressure before it is accelerated to speed $\triangle v$.   For transonic and subsonic flows, the tidal drag is much less.

\end{document}